\documentclass{PoS}
\usepackage{epsfig}
\usepackage{amsmath}
\usepackage{amssymb}

\PoS{PoS(LAT2005)085}

\title{Effects of partial quenching and staggered fermions on the scalar correlator }

\ShortTitle{Effects of partial quenching and staggered fermions on the scalar correlator }

\author{\speaker{Sasa Prelovsek}\\
    University of Ljubljana and Institute Jozef Stefan, Ljubljana, Slovenia\\
        E-mail: \email{sasa.prelovsek@ijs.si}}

\abstract{
We determine the mass of the lightest  
$\bar qq$ scalar meson with $I=1$  
using the simulation with two dynamical Domain Wall Fermions. 
The conventional exponential fit of the scalar correlator is justified in 
this case giving the mass $1.58\pm 0.34$ GeV. In general the 
scalar correlator receives 
also the bubble contribution, which is the intermediate state 
with two pseudoscalar mesons. This contribution is sizable at 
light quark masses and has to be incorporated in the fit of the 
scalar correlator in order to extract the scalar meson mass. We provide 
predictions for the bubble contribution in Partially Quenched ChPT, 
Staggered ChPT and ChPT for Mixed quark actions. We find that 
the bubble contribution is significantly affected by the 
unphysical approximations that are employed in simulations. 
It can  render the negative sign and 
unphysical effective mass $2M_\pi$ in the scalar correlator with $I=1$.  
}

\FullConference{XXIIIrd International Symposium on Lattice Field Theory\\
		 25-30 July 2005\\
		 Trinity College, Dublin, Ireland}

\begin{document}

\section{Introduction}

The Nature of the lightest observed scalar resonances is 
not revealed yet. 
There are two experimentally well established 
scalar resonances $a_0(980)$ and $a_0(1450)$ with isospin $I=1$ below 
$2$ GeV. It is still not clear which one of the two is the lightest 
$\bar du$ scalar state. This raises a further question whether 
$a_0(980)$ is perhaps a tetraquark \cite{tetraquark} if $a_0(1450)$ 
turns out to be the lightest $\bar du$ scalar state.  

\begin{figure}[htb!]
\begin{center}
\epsfig{file=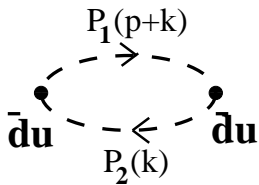,height=3cm}
\end{center}

\vspace{-0.8cm}

\caption{ \small The bubble contribution to the scalar correlator. 
Here $P_1$ and $P_2$ denote pseudoscalar mesons in the relevant 
version of ChPT. }\label{fig.bubble}
\end{figure}

The first issue could be settled with a determination of the lightest 
$\bar du$ scalar mass on the lattice. For this purpose the lattice simulations evaluate the scalar correlator 
\begin{equation}
\label{cor}
C(t)=\sum _{\vec x}\langle 0|\bar d (\vec x,t) u(\vec x,t)  
~\bar u(\vec 0,0)d(\vec 0,0)|0\rangle=Ae^{-m_{a0}t}+B(t)+\cdots~.
\end{equation}
If $a_0$ is the lightest state with $I=1$ and $J^P=0^+$, 
the correlator (\ref{cor}) drops as $e^{-m_{a0}t}$ at large $t$ 
and determination of $m_{a0}$ is straight forward. 
Multi-hadron states with $J^P=0^+$ and $I=1$ 
also propagate between the source and the sink 
and they often 
shadow the interesting part $e^{-m_{a0}t}$ in the correlator (\ref{cor}). 
The most important multi-hadron state  is 
 {\it the intermediate state with two pseudoscalars}, we call 
it {\it the bubble contribution} $B(t)$ and display it in 
Fig. \ref{fig.bubble}. 
In proper three flavor QCD the 
two-pseudoscalar states are $\pi\eta, K\bar K,\pi\eta^\prime$, 
while $\pi\pi$ is not allowed by Bose symmetry and
 conservation of $J^P$ and $I^G$.   
 In  two flavor QCD the only state  
$\pi\eta^\prime$ is relatively heavy and therefore not so disturbing. 
The scalar correlator (\ref{cor}) is dominated by $B(t)$ 
in the simulations with light quark masses if  $M_{P1}+M_{P2}<m_{a0}$ 
which opens the  decay channel $a_0\to P_1P_2$. 
So the bubble contribution has to be incorporated in the 
fit of the scalar correlator (\ref{cor}) in order to extract $m_{a0}$.  

In addition to the physical intermediate states $P_1P_2$, the bubble 
contribution incorporates also the effects of the 
unphysical approximations employed in the simulations. 
This allows the extraction of $m_{a0}$ via (\ref{cor}) even in this case. 
The unphysical effects 
were first attributed to the bubble contribution in case of quenched QCD 
\cite{bardeen}, where it describes  
the magnitude and the negative sign of the 
lattice correlator  well \cite{bardeen,sasa_quenched}. 

\vspace{-0.5cm}

Here we present the analytical predictions for $B(t)$ in
 partially quenched simulations \cite{sasa_pq} and  in 
simulations with staggered fermions or mixed-quark actions \cite{sasa_s}. 
The results are obtained  at the lowest order in  
the appropriate version of Chiral Perturbation Theory (ChPT) and 
apply for the point-point correlators. 

We demonstrate the method of determining $m_{a0}$    
in a simulation with two dynamical Domain Wall Fermions (DWF) 
\cite{sasa_pq}. 
The resulting scalar meson masses from the dynamical and partially quenched 
correlators  agree. 

The simulation with chiral (DWF) sea and valence quarks 
is presented in Section 2. Section 3 considers the simulations with 
staggered sea and valence quarks. The simulations with 
staggered sea  and chiral valence are considered in Section 4. We conclude  
in Section 5.

\section{Simulation with two dynamical Domain Wall quarks}

{\bf Dynamical correlator with $m_{val}=m_{sea}$}

\vspace{0.1cm}

\begin{figure}[htb!]
\begin{center}
\epsfig{file=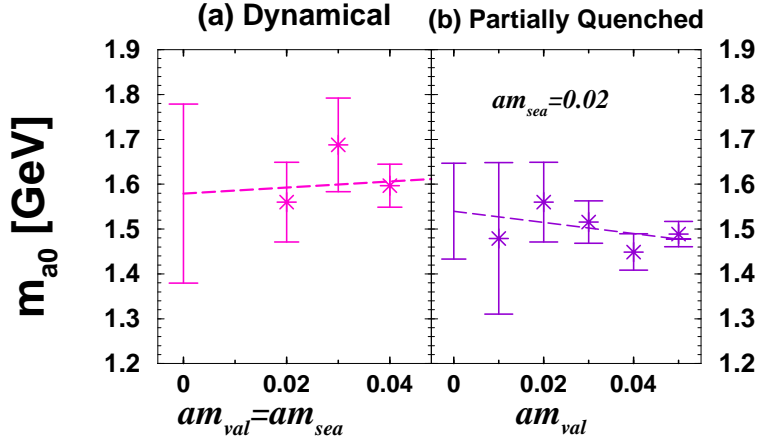,height=6cm}
\end{center}

\vspace{-0.8cm}

\caption{\small Mass of $a_0$ from a simulation with two dynamical 
Domain Wall quarks, $a^{-1}\simeq 1.7$ GeV and $V=16^3\times 32$:
(a) $m_{a0}$ from dynamical scalar correlator 
with $am_{val}=am_{sea}$ using exponential fit; (b) $m_{a0}$ 
from partially quenched 
scalar correlator with  $am_{val}\! \not{\! =}am_{sea}=0.02$ using fit 
(1.1) with $B^{PQChPT}$ (2.2).}\label{fig.m_dwf}
\end{figure}

The simulation with two  
Domain Wall sea quarks and Domain Wall valence quarks  \cite{RBC_dyn} 
does not suffer 
from unphysical effects in contrast to the simulations 
that will be considered later. The conventional exponential fit 
$C(t)=Ae^{-m_{a0}t}$ (\ref{cor}) is justified in two-flavor QCD  
since the only intermediate state $\pi\eta^\prime$ is relatively heavy 
and $B(t)$ is small  \cite{sasa_pq}. The resulting $m_{a0}$ for 
three different quark masses is presented in Fig. \ref{fig.m_dwf}a,
 while the linear extrapolation to the chiral limit gives 
\begin{equation}
\label{m_dyn}
m_{a0}^{dyn}=1.58\pm 0.34\ {\rm GeV}~.
\end{equation}     
This mass has  large error-bar, but it is 
definitely above $1$ GeV and favors $a_0(1450)$ as the lightest $\bar d u$ 
scalar meson. 

{\bf Partially Quenched correlator with $m_{val}\not =m_{sea}$}

\vspace{0.1cm}

\begin{figure}[htb!]
\begin{center}
\epsfig{file=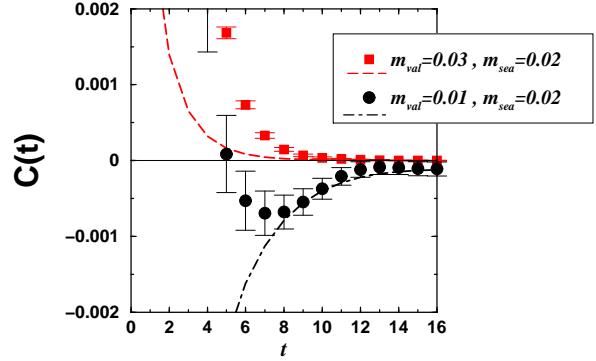,height=5cm}
\end{center}

\vspace{-0.8cm}

\caption{ \small The symbols present the 
partially quenched point-point lattice correlators from a simulation 
with two dynamical Domain Wall quarks. The lines present 
the corresponding $B^{PQChPT}(t)$ (2.2).}\label{fig.pq_compare}
\end{figure}

The partially quenched scalar correlator in Fig. \ref{fig.pq_compare} 
is positive for $m_{val}>m_{sea}$, whereas it is negative for 
$m_{val}<m_{sea}$ \cite{sasa_pq}. The striking effect of partial 
quenching is attributed to the bubble diagram in Fig. \ref{fig.bubble}. 
In order to compute the bubble diagram we need the coupling of 
the point scalar current to two pseudoscalars and the pseudoscalar 
propagators 
from Partially Quenched ChPT. The coupling is obtained 
from $\bar du=-\partial{\cal L}^{ChPT}/\partial {\cal M}_{du}$ 
and is equal to $B_0=M_\pi^2/(2m_q)$ in any 
version of $ChPT$ at the lowest order. The bubble contribution 
for $m_0\to\infty$ is $B(t)=F.T.[B(p)]_{\vec p=0}$ with 
\cite{sasa_pq} 
\begin{equation}
\label{Bpqchpt}
B^{PQChPT}(p)=2B_0^2 \! \sum_k\Biggl\{\frac{1}{(k+p)^2+M_{val,sea}^2}~ \frac{1}{k^2+M_{val,sea}^2}-\frac{1}{(k+p)^2+M_{val,val}^2}~\frac{k^2+M_{sea,sea}^2}{(k^2+M_{val,val}^2)^2}\Biggl\}~.
\end{equation}
It does not contain any free parameters, since pseudoscalar masses $M$ 
and $B_0=M_\pi^2/(2m_q)$ are determined from the 
lattice data. The sum over the loop momenta $k$ 
is performed over the allowed discrete momenta on the lattice. 
Fig. \ref{fig.pq_compare} shows that the bubble contribution is positive for 
$m_{val}>m_{sea}$ and negative for $m_{val}<m_{sea}$ like the lattice data.  
The negativity of the scalar correlator for $m_{val}<m_{sea}$ and large $t$ 
is described well by the bubble contribution. Having understood 
the source of the unphysical effect of partial quenching, we extract 
$m_{a0}$ by fitting the partially quenched scalar correlator to 
(\ref{cor}) and display it in Fig. \ref{fig.m_dwf}a. 
The linear extrapolation 
to the chiral limit gives $m_{a0}^{PQ}=1.51\pm 0.19~$GeV in agreement  
with the dynamical result (\ref{m_dyn}) and with smaller error.  

\section{Simulations with staggered sea and valence quarks}

The scalar correlator was simulated using $2+1$ staggered sea quarks and 
staggered valence quarks  by MILC \cite{milc} and UKQCD \cite{irving}. 
The authors of \cite{milc,irving} were 
surprised to find the effective mass significantly below $M_{\pi}+M_\eta$, 
although $\pi\eta$ is the lightest state with $I=1$ for  
light $u/d$ quarks in proper QCD.

\begin{figure}[htb!]
\begin{center}
\epsfig{file=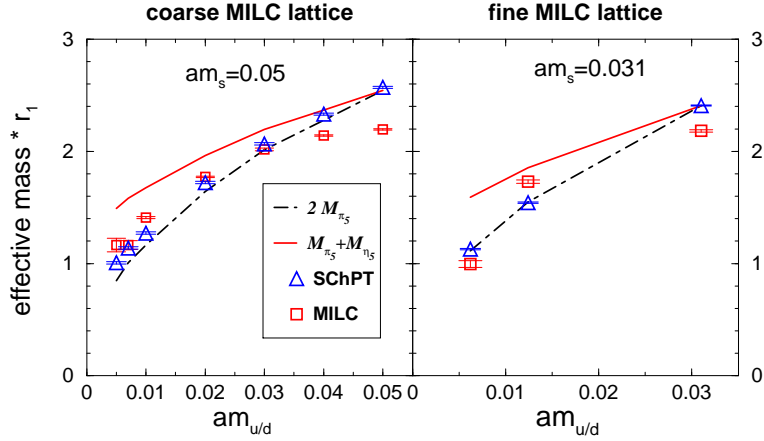,height=6cm}
\end{center}

\vspace{-0.8cm}

\caption{ \small The squares represent the  
effective mass of the scalar correlator from MILC 
staggered simulation with $2+1$ dynamical quarks. The triangles 
present the effective mass of $B^{SChPT}(t)$ (3.1) at 
$t=20$ for coarse MILC lattice and at $t=30$ for fine MILC lattice. 
Mass is plotted in units of $r_1\simeq 1.6$ GeV$^{-1}$ like in 
\cite{milc}.}\label{fig.schpt_eff}
\end{figure}

This can be attributed to the taste breaking effects which 
enter the scalar correlator via the bubble contribution \cite{sasa_s}
in the Staggered ChPT of \cite{schpt}
\begin{align}
\label{Bschpt}
B^{SChPT}(p)=B_0^2 \! \sum_k\Biggl\{
-4\biggl[&\frac{1}{(k+p)^2+M_{U_I}^2}~ \frac{1}{3}~
\frac{(k^2+M_{S_I}^2)}{(k^2+M_{U_I}^2)(k^2+\tfrac{1}{3}M_{U_I}^2+\tfrac{2}{3}M_{S_I}^2)}\\
+~&\frac{1}{(k+p)^2+M_{U_V}^2}~a^2\delta_V~\frac{(k^2+M_{S_V}^2)}{(k^2+M_{U_V}^2)(k^2+M_{\eta_V}^2)(k^2+M_{\eta^\prime_V}^2)}~+~(V\to A)\biggr]\nonumber\\
+~\frac{1}{16}\sum_{b=1}^{16}& \biggl[2~ \frac{1}{(k+p)^2+M_{U_b}^2}~ \frac{1}{k^2+M_{U_b}^2}+\frac{1}{(k+p)^2+M_{us_{b}}^2}~ \frac{1}{k^2+M_{us_{b}}^2}\biggr]\Biggl\}~.\nonumber
\end{align}
with the notation from \cite{schpt}. The forth-root trick is incorporated in 
(\ref{Bschpt}) by weighting the diagrams with quark loops by  factor $1/4$.   
The prediction has no free parameters since the pseudoscalar masses 
of various tastes \cite{milc} and hairpin parameters 
$a^2\delta^\prime_{V,A}$ \cite{milc3} have been determined by MILC.  
In the continuum limit 
the pseudoscalars 
of various tastes are degenerate and $a^2\delta^\prime_{V,A}\to 0$, 
so the lightest intermediate state in (\ref{Bschpt}) 
is  $\pi\eta$. The taste breaking 
at finite $a$ makes possible the  intermediate state with mass  
$2M_\pi$ and is responsible that the effective mass of 
$B^{SChPT}(t)$  is close to $2M_\pi$, as shown in Fig. 
\ref{fig.schpt_eff}. The effective mass of $B^{SChPT}(t)$ 
qualitatively agrees with the effective mass obtained by MILC \cite{milc} and 
UKQCD \cite{irving} indicating that the lattice correlators are dominated by 
two-pseudoscalar intermediate states and are significantly affected by the taste breaking. 

\vspace{-0.2cm}

\section{Simulations with staggered sea quarks and chiral valence quarks}

The simulations with chiral valence quarks 
on the available staggered MILC configurations \cite{mixed_lhp,mixed_ukqcd} 
present an appealing possibility. However there 
is no unique recipe how to match 
valence and sea quark masses at finite $a$  
and some features of partial quenching always remain.

\begin{figure}[htb!]
\begin{center}
\epsfig{file=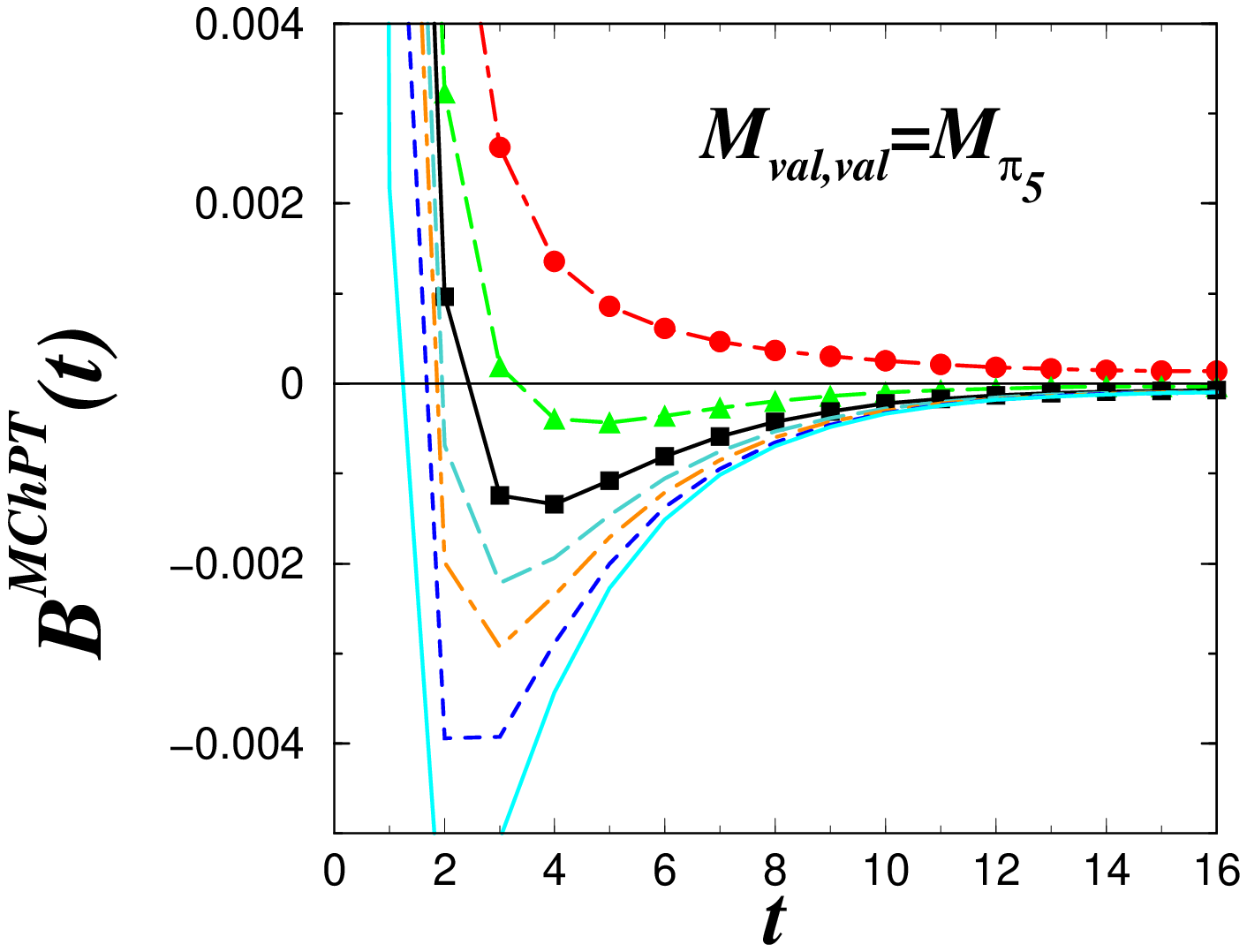,height=5cm}
$\quad$
\epsfig{file=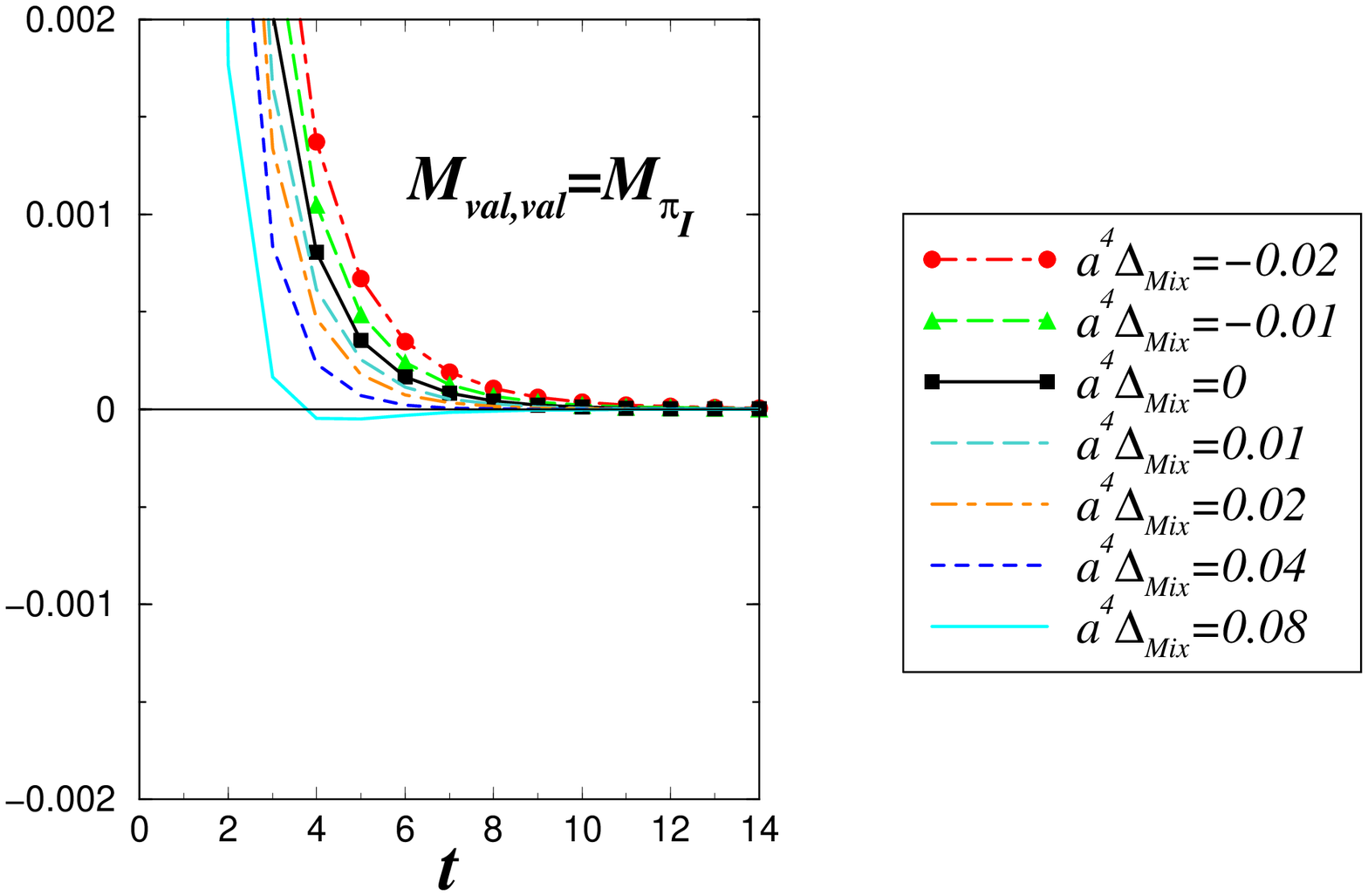,height=5cm}
\end{center}

\vspace{-0.8cm}

\caption{ \small The bubble contribution $B^{MChPT}(t)$ (4.1) 
for simulation 
with chiral fermions on coarse MILC configurations 
with sea quark masses 
$am_{u/d}=0.01$ and $am_{s}=0.05$. It is plotted for two choices 
of tuning $m_{val}$ with $m_{sea}$  and it depends on the 
value of $a^2\Delta_{Mix}~$, which is varied in the reasonable range 
\cite{sasa_s}.}\label{fig.mchpt}
\end{figure}

The effect of mixed quark actions in simulations with $2+1$ 
staggered sea quarks can be attributed to the bubble 
contribution \cite{sasa_s} within Mixed ChPT of \cite{mchpt}\footnote{Our result for $B^{MChPT}(t)$ agrees in the limit $t\to\infty$, $a^2\Delta_{Mix}\to 0$ and $m_u=m_d=m_s$ with the result of \cite{mchpt_taku}, 
who study  mixed quark actions but not the staggered sea.}
\begin{align}
\label{Bmchpt}
B^{MChPT}(p)&=B_0^2 \! \sum_k\Biggl\{-\frac{4}{3}~\frac{1}{(k+p)^2+M_{val,val}^2}~\frac{1}{(k^2+M_{val,val}^2)^2}~
\frac{(k^2+M_{U_I}^2)(k^2+M_{S_I}^2)}{k^2+\tfrac{1}{3}M_{U_I}^2+\tfrac{2}{3}M_{S_I}^2}\\
&\ \ \ \ \ \ \ \ \ \ \ \ +2~ \frac{1}{(k+p)^2+M_{val,u}^2}~ \frac{1}{k^2+M_{val,u}^2}+\frac{1}{(k+p)^2+M_{val,s}^2}~ \frac{1}{k^2+M_{val,s}^2}\Biggl\}~.\nonumber
\end{align}
The only unknown parameter   
$a^2\Delta_{Mix}\!\equiv\! M_{val,sea}^2-B_0(m_{sea}+m_{val})$, {\small $sea=u,s~$},  
gives the taste breaking in the pion mass composed of 
one valence and one sea quark \cite{mchpt}, while the other 
input parameters have been determined in \cite{milc,mixed_lhp}. 
Fig. \ref{fig.mchpt} shows that the scalar correlator can 
be negative for light $u/d$ quarks if valence and sea quark masses 
are tuned by matching $M_{val,val}=M_{\pi_5}$ 
(used by LHP \cite{mixed_lhp}),  
while it is 
positive if  $M_{val,val}=M_{\pi_I}$ are matched \cite{sasa_s}.  
Comparison of the point-point lattice correlators and $B^{MChPT}$ 
in case of matching $M_{val,val}\!\!=\!M_{\pi_5}$  offers a possibility 
to determine $a^2\Delta_{Mix}$.  

\vspace{-0.2cm}

\section{Conclusions}

\vspace{-0.2cm}

We determined the mass of the lightest  
$\bar qq$ scalar meson with $I=1$  
using the simulation with two dynamical Domain Wall Fermions. 
The exponential fit of the scalar correlator is justified in 
this case giving the mass $1.58\pm 0.34$ GeV.
In general the scalar correlator receives 
also a sizable bubble contribution $B(t)$, which is the intermediate state 
with two pseudoscalar mesons. We provide analytical predictions for 
$B(t)$ for simulations that use (partial) quenching, staggered fermions 
or mixed quark actions. Our predictions for $B(t)$ 
within relevant versions of ChPT 
are free of unknown parameters, expect for a parameter $a^2\Delta_{Mix}$ in case of mixed quark actions,  
which could be determined from the scalar correlator. We find that $B(t)$ 
is sizable at small quark masses and  is significantly affected by the 
unphysical approximations that are employed in simulations. 
It can  render the negative sign and 
unphysical effective mass $2M_\pi$ in the scalar correlator. 
Our predictions for the bubble contribution will be needed in order to 
extract the scalar meson mass from the scalar correlator 
in the future simulations.  

%\vspace{-0.3cm}

\end{document}